\documentclass[prl,groupedaddress,showpacs,twocolumn]{revtex4}
\usepackage[pdftex]{graphicx}
\usepackage{amssymb}
\newcommand{\ped}[1]{\ensuremath{_{\rm #1}}}

\begin{document}
\title{Temperature-dependent spin resonance energy in iron pnictides and multiband $s \pm$ Eliashberg theory}
\author{G.A. Ummarino}
\email{E-mail:giovanni.ummarino@infm.polito.it}
\affiliation{Dipartimento di Fisica and CNISM, Politecnico di
Torino, Corso Duca degli Abruzzi 24, 10129 Torino, Italy}

\begin{abstract}
The phenomenology of iron-pnictides superconductors can be explained
in the framework of a three bands s$\pm$ wave Eliashberg theory with
only two free parameters plus a \emph{feedback effect} i.e. the
effect of the condensate on the antiferromagnetic spin fluctuactions
responsible of the superconductivity in these compounds. I have
examined the experimental data of four materials:
$\mathrm{LaFeAsO_{1-x}F_{x}}$, $\mathrm{SmFeAsO_{1-x}F_{x}}$,
$\mathrm{Ba_{1-x}K_{x}Fe_{2}As_{2}}$, and
$\mathrm{Ba(Fe_{x}Co_{1-x})_{2}As_{2}}$ and I have found that it is
possible to reproduce the experimental critical temperature and gap
values in a moderate strong-coupling regime: $\lambda_{tot}\approx
1.7-2.0$.
\end{abstract}


\maketitle

The new class of Fe-based compounds \cite{Kamihara_La, Ren_Sm,
Wang_Gd} just as the cuprates \cite{Chubukov} and the heavy fermions
\cite{HF} have all some similar caracteristics. For example the high
values of rate $2\Delta/T_{c}$ or the presence of the pseudogap
\cite{Chubukov,HFAR,Gonnelli_La}. For all three class of material it
is proposed the superconductivity to be mediated by
antiferromagnetic spin fluctuactions
\cite{Chubukov,HFTH,Mazin_PhysC_SI}. The most obvious difference is
that almost all the iron compounds present a multiband behavior
while in HTCS and in heavy fermions this was detected only in some
particular cases. The multi-band nature of Fe-based superconductors
may give rise to a multi-gap scenario \cite{Tes} that is indeed
emerging from many different experimental data with evidence for
rather high gap ratios, $\approx 2-3$ \cite{PhysC_Fe}. In this
regard neither a three-band BCS model \cite{Mazin_PhysC_SI,
Benfatto,Kuchinskii} nor a four-band Eliashberg model \cite{Ema}
with small values of the coupling constants and large boson energies
are adequate: the former can only account for the gap ratio and
$T_c$ but not for the exact experimental gap values and the latter
provides a calculated critical temperature larger than the
experimental one. The high experimental value of the larger gap
suggests that high values of the coupling constants might be
necessary to explain the experimental data within a three-band model
\cite{Umma1,Umma2}: one has therefore to employ the Eliashberg
theory for strong coupling superconductors \cite{Umma1,Umma2}. In my
early works \cite{Umma1,Umma2} I found that a three-band Eliashberg
model allows to reproduce various experimental data, this suggests
that these compounds can represent a case of dominant negative
interband-channel superconductivity ($s \pm$ wave symmetry) with
small typical boson energies ($\approx 10$ meV) but too high values
of the electron-boson coupling constants ($1.9 \leq\lambda_{tot}\leq
5.9$). The way for solve this problem is suggested by experimental
measurement of Inosov and coworkers \cite{Inosov}: they find that
the temperature evolution of the spin resonance energy follows the
superconducting energy gap and this should indicate a \emph{feedback
effect} \cite{Chubukov,FBB,FB} of the condensate on the spin
fluctuactions. I assume that this is the starting point of my
argumentation. The procedure is as follows: first of all I choose
the experimental low temperature spin resonance as representative
boson energy and I fix the two remaining free parameters to
reproduce the exact experimental gap values. then, with the same
parameters, I calculate the critical temperature $T^{*}_{c}$. I find
always $T^{*}_{c}\gg T^{exp}_{c}$ where $T^{exp}_{c}$ is the
experimental critical temperature. In the successive step I use the
same input parameters utilized before except for the electron-boson
spectral functions that have an energy peak with the same
temperature dependence of the superconductive gap. Of course at
$T=T^{*}_{c}$ the energy peak is equal to zero while at $T=0$ K the
new spectral functions are equal to old ones. In this way, taking
into account the \emph{feedback effect} of the condensate
\cite{Chubukov,FBB,FB} on the antiferromagnetic spin fluctuactions I
could explain the experimental data (the gap values and the critical
temperature) in a model which has only two free parameters in a
moderate strong coupling regime ($\lambda_{tot}\approx 1.7-2$).

I choose four representative cases (three hole type and one electron
type): $\mathrm{LaFeAsO_{1-x}F_{x}}$, $\mathrm{SmFeAsO_{1-x}F_{x}}$,
$\mathrm{Ba_{1-x}K_{x}Fe_{2}As_{2}}$, and
$\mathrm{Ba(Fe_{x}Co_{1-x})_{2}As_{2}}$. The electronic structure of
the compounds hole type can be approximately described by a
three-band model \cite{Mazin_PhysC_SI} with two hole bands
(indicated in the following as bands 1 and 2) and one equivalent
electron band (3) \cite{Umma1,Umma2} while for one electron type
with one hole band (indicated in the following as band 1) and two
equivalent electron bands (2 and 3) \cite{Tortello2010}. In the hole
type case the $s$-wave order parameters of the hole bands
$\Delta_{1}$ and $\Delta_{2}$ have opposite sign compared to
electron band one, $\Delta_{3}$ \cite{Mazin_spm} while, in the
electron type case, $\Delta_{1}$ has opposite sign compared to two
electron bands ones, $\Delta_{2}$ and $\Delta_{3}$
\cite{Tortello2010} . In such systems, intraband coupling could be
provided by phonons (\textit{ph}), and interband coupling by
antiferromagnetic spin fluctuations (\textit{sf}) \cite{Mazin_spm}.
I summarize the experimental data relative to the four considered
cases:

1) the compound $\mathrm{LaFeAsO_{0.9}F_{0.1}}$ (LaFeAsOF)with
$T^{A}_{c}=28.6$ K where point-contact spectroscopy measurements
gave $\Delta_{1}(0)\approx 8.0 $ meV and $\Delta_{2}(0)\approx 2.8 $
meV \cite{Gonnelli_La};

2) $\mathrm{Ba_{0.6}K_{0.4}Fe_{2}As_{2}}$ (BaKFeAs) with $T_{c}=37$
K where ARPES measurements gave $\Delta_{1}(0)=12.1 \pm 1.5$ meV,
$\Delta_{2}(0)=5.2 \pm 1.0$ meV and $\Delta_{3}(0) =12.8 \pm 1.4$
meV \cite{Ding_ARPES_BaKFeAs};

3) the compound $\mathrm{SmFeAsO_{0.8}F_{0.2}}$ (SmFeAsOF) with
$T^{A}_{c}=52$ ($T^{bulk}_{c}=53$ K) K where, according to
point-contact spectroscopy measurements, $\Delta_{1}(0)=18\pm 3$ meV
and $\Delta_{2}(0)=6.2\pm 0.5 $ meV \cite{Daghero_Sm};

4) the compound $\mathrm{Ba(Fe_{x}Co_{1-x})_{2}As_{2}}$ (BaFeCoAs)
with $T^{A}_{c}=22.6$ K ($T^{bulk}_{c}=24.5$ K) where, according to
point-contact spectroscopy measurements, $\Delta_{1}(0)=4.1\pm 0.4$
meV and $\Delta_{2}(0)=9.2\pm 1.0$ meV \cite{Tortello2010}.

$T^{A}_{c}$ is the critical temperature obtained by Andreev
reflection measurements and $T^{bulk}_{c}$ is the critical
temperature extracted by transport measurements. Note that only in
the case of ARPES the gaps are associated to the relevant band since
point-contact spectroscopy measurements generally gives only two
gaps, the larger one has been arbitrarily indicated as $\Delta_1$
supposing that $\Delta_1 \sim |\Delta_3|$.
\begin{table}[!]
\vspace{-3mm}
\begin{tabular}{|c|c|c|c|c|c|c|}
\hline
                  & $\lambda_{tot}$ & $\lambda^{old}_{tot}$  & $\lambda_{12/21}$  & $\lambda_{13/31}$ &  $\lambda_{23/32}$ &$\Omega_0$ (meV)\\
\hline

                   &1.87&       & 0.76/0.85 &  1.21/5.44& 0.00/0.00 &9.04    \\
  BaFeCoAs         &2.83&  1.93 & 0.91/1.02 &  2.08/9.35& 0.00/0.00 &9.04      \\
\hline
                    &1.75&      & 0.00/0.00 &  2.11/1.91& 0.40/0.21 &11.44     \\
  LaFeAsOF          &2.38& 2.53 & 0.00/0.00 &  2.93/2.66& 0.46/0.24 & 11.44       \\
\hline
                   & 2.04&        & 0.00/0.00 &  2.27/2.27& 0.56/0.28 &14.80  \\
  BaKFeAs          & 2.84& 3.87   & 0.00/0.00 &  3.21/3.21& 0.67/0.34 &14.80 \\
\hline
                   & 1.72&      & 0.00/0.00 &  1.55/3.88& 0.42/0.84 &20.80 \\
  SmFeAsOF         & 2.39& 5.90 & 0.00/0.00 &  2.23/5.58& 0.49/0.98 &20.80 \\
\hline
\end{tabular}
\caption{The values of $\Omega_0$ and $\lambda_{ij}$, that allow
reproducing the experimental gap values, are shown. $\lambda_{tot}$
is compared with $\lambda^{old}_{tot}$ that is the value determined
in the previous works \cite{Umma1,Umma2,Tortello2010}. In the first
arrows the \textit{sf} spectral functions used have usual shape
while in the second ones have Lorentzian shape.}\label{table:1}
\end{table}
\begin{table}[!]
\vspace{-3mm}
\begin{tabular}{|c|c|c|c|c|c|}
\hline
& $\Delta_{1}(meV)$ & $\Delta_{2}(meV)$   & $\Delta_{3}(meV)$  & $T_{c}(K)$ &  $T^{*}_{c}(K)$ \\
  \hline
               & 6.63 & -4.07 &  -9.18&  26.07&  33.00               \\
  BaFeCoAs     & 7.02 & -4.12 &  -9.18&  23.73&  28.95               \\
             \hline
               & 8.01 &  2.82 & -7.75  &  29.37&  37.22               \\
  LaFeAsOF     & 8.01 &  2.77 & -7.71  &  26.86&  31.81               \\
\hline
               & 12.04 &  5.20 & -12.00  &  43.66&  55.26               \\
  BaKFeAs      & 12.04 &  5.24 & -11.91  &  38.33&  46.18               \\
\hline
               & 14.86 &  6.15 & -18.11  &  58.53&  74.13             \\
  SmFeAsOF     & 15.51 &  6.15 & -18.00  &  52.80&  63.82             \\
\hline
\end{tabular}
\caption{The calculated values of the gaps and of the two critical
temperature with and without \emph{feedback effect}. In the first
arrows the \textit{sf} spectral functions used have usual shape
while in the second ones have Lorentzian shape.}\label{table:3}
\end{table}

To obtain the gaps and the critical temperature within the s$\pm$
wave, three-band Eliashberg equations \cite{Eliashberg} one has to
solve six coupled equations for the gaps $\Delta_{i}(i\omega_{n})$
and the renormalization functions $Z_{i}(i\omega_{n})$, where $i$ is
a band index (that ranges between $1$ and $3$) and $\omega_{n}$ are
the Matsubara frequencies. If one neglects for simplicity the effect
of magnetic and non-magnetic impurities, the imaginary-axis
equations \cite{Umma1,Umma2} are:
\begin{equation}
\omega_{n}Z_{i}(i\omega_{n})=\omega_{n}+ \pi
T\sum_{m,j}\Lambda^{Z}_{ij}(i\omega_{n},i\omega_{m})N^{Z}_{j}(i\omega_{m})\label{eq:EE1}
\end{equation}
\begin{eqnarray}
Z_{i}(i\omega_{n})\Delta_{i}(i\omega_{n})&=& \pi
T\sum_{m,j}[\Lambda^{\Delta}_{ij}(i\omega_{n},i\omega_{m}) \label{eq:EE2}\\
& &
-\mu^{*}_{ij}(\omega_{c})]\Theta(\omega_{c}-|\omega_{m}|)N^{\Delta}_{j}(i\omega_{m})\vspace{5mm}\nonumber
\end{eqnarray}
where
$\Lambda^{Z}_{ij}(i\omega_{n},i\omega_{m})=\Lambda^{ph}_{ij}(i\omega_{n},i\omega_{m})+\Lambda^{sf}_{ij}(i\omega_{n},i\omega_{m})$,
$\Lambda^{\Delta}_{ij}(i\omega_{n},i\omega_{m})=\Lambda^{ph}_{ij}(i\omega_{n},i\omega_{m})-\Lambda^{sf}_{ij}(i\omega_{n},i\omega_{m})$.
$\Theta$ is the Heaviside function and $\omega_{c}$ is a cutoff
energy. In particular,
$\Lambda^{ph,sf}_{ij}(i\omega_{n},i\omega_{m})=2
\int_{0}^{+\infty}d\Omega \Omega
\alpha^{2}_{ij}F^{ph,sf}(\Omega)/[(\omega_{n}-\omega_{m})^{2}+\Omega^{2}]$.
$\mu^{*}_{ij}(\omega\ped{c})$ are the elements of the $3\times 3$
Coulomb pseudopotential matrix. Finally,
$N^{\Delta}_{j}(i\omega_{m})=\Delta_{j}(i\omega_{m})/
{\sqrt{\omega^{2}_{m}+\Delta^{2}_{j}(i\omega_{m})}}$ and
$N^{Z}_{j}(i\omega_{m})=\omega_{m}/{\sqrt{\omega^{2}_{m}+\Delta^{2}_{j}(i\omega_{m})}}$.
The electron-boson coupling constants are defined as
$\lambda^{ph,sf}_{ij}=2\int_{0}^{+\infty}d\Omega\frac{\alpha^{2}_{ij}F^{ph,sf}(\Omega)}{\Omega}$.

The solution of eqs. \ref{eq:EE1} and \ref{eq:EE2} requires a huge
number of input parameters (18 functions and 9 constants); however,
some of these parameters are related one to another, some can be
extracted from experiments and some can be fixed by suitable
approximations. As shown in Ref. \cite{Mazin_spm}, in the case of
pnictides we can assume that: i) the total electron-phonon coupling
constant is small \cite{Boeri2}; ii) phonons mainly provide
intraband coupling; iii) spin fluctuations mainly provide interband
coupling. To account for these assumptions in the simplest way, I
will take: $\lambda^{ph}_{ii}=\lambda^{ph}_{ij}=0.$ (upper limit of
the phonon coupling \cite{Boeri2} $\approx0.35$),
$\lambda^{sf}_{ii}=0$ (only interband \textit{sf} coupling) and
$\mu^{*}_{ii}(\omega\ped{c})=\mu^{*}_{ij}(\omega\ped{c})=0$
\cite{Umma1}. Within these approximations, the electron-boson
coupling-constant matrix $\lambda_{ij}$ becomes:
\cite{Mazin_PhysC_SI, Umma1,Tortello2010}:
\begin{equation}
\vspace{2mm} %
\lambda_{ij}= \left (
\begin{array}{ccc}
  0                 &          \lambda_{12}                  &               \lambda_{13}            \\
  \lambda_{21}=\lambda_{12}\nu_{12}                &               0               &               \lambda_{23}            \\
  \lambda_{31}=\lambda_{13}\nu_{13} & \lambda_{32}=\lambda_{23}\nu_{23}    & 0 \\
\end{array}
\right ) \label{eq:matrix}
\end{equation}
where $\nu_{ij}=N_{i}(0)/N_{j}(0)$ and $N_{i}(0)$ is the normal
density of states at the Fermi level for the $i$-th band. In the
hole case it is $\lambda_{21}=\lambda_{12}=0$ while in the electron
case $\lambda_{23}=\lambda_{32}=0$.
In the numerical simulations I used the standard form for the
antiferromagnetic spin fluctuaction \cite{dolghi}:
$\alpha^{2}_{ij}F^{sp}(\Omega)=B_{ij}\Omega\cdot
\Omega_{ij}\cdot\Theta(\Omega_{max}-\Omega)/(\Omega^{2}+\Omega^{2}_{ij})$
where $B_{ij}$ are the normalization constants necessary to obtain
the proper values of $\lambda_{ij}$ while $\Omega_{ij}$ are the peak
energies. In all the calculations I always set
$\Omega_{ij}=\Omega_{0}$. The maximum \textit{sf} energy is
$\Omega_{max}=10 \Omega_{0}$, the cut-off energy is $\omega_{c}=30
\Omega_{0}$ and the maximum quasiparticle energy is $\omega_{max}=40
\Omega_{0}$. As typical \textit{sf} energy $\Omega_{0}$ I use the
spin resonance energy that have been measured and I assume correct
for all compounds examined the relation $\Omega_{0}=(2/5)T_{c}$
available in literature \cite{Paglione}. Bandstructure calculations
provide information about the factors $\nu_{ij}$ that enter in the
definition of $\lambda_{ij}$ (eq. 3). In the case of
$\mathrm{LaFeAsO_{0.9}F_{0.1}}$ I know that $\nu_{13}=0.91$ and
$\nu_{23}=0.53$ \cite{Mazin4}, in
$\mathrm{Ba_{0.6}K_{0.4}Fe_{2}As_{2}}$ $\nu_{13}=1$ and $\nu_{23}=2$
\cite{Mazin_PhysC_SI}, in $\mathrm{SmFeAsO_{0.8}F_{0.2}}$
$\nu_{13}=0.4$ and $\nu_{23}=0.5$ \cite{Mazin4} and in
$\mathrm{Ba(Fe_{x}Co_{1-x})_{2}As_{2}}$ $\nu_{12}=1.12$ and
$\nu_{13}=4.50$ \cite{Mazin4}.
\begin{figure}[!]
\begin{center}
\includegraphics[keepaspectratio, width=\columnwidth]{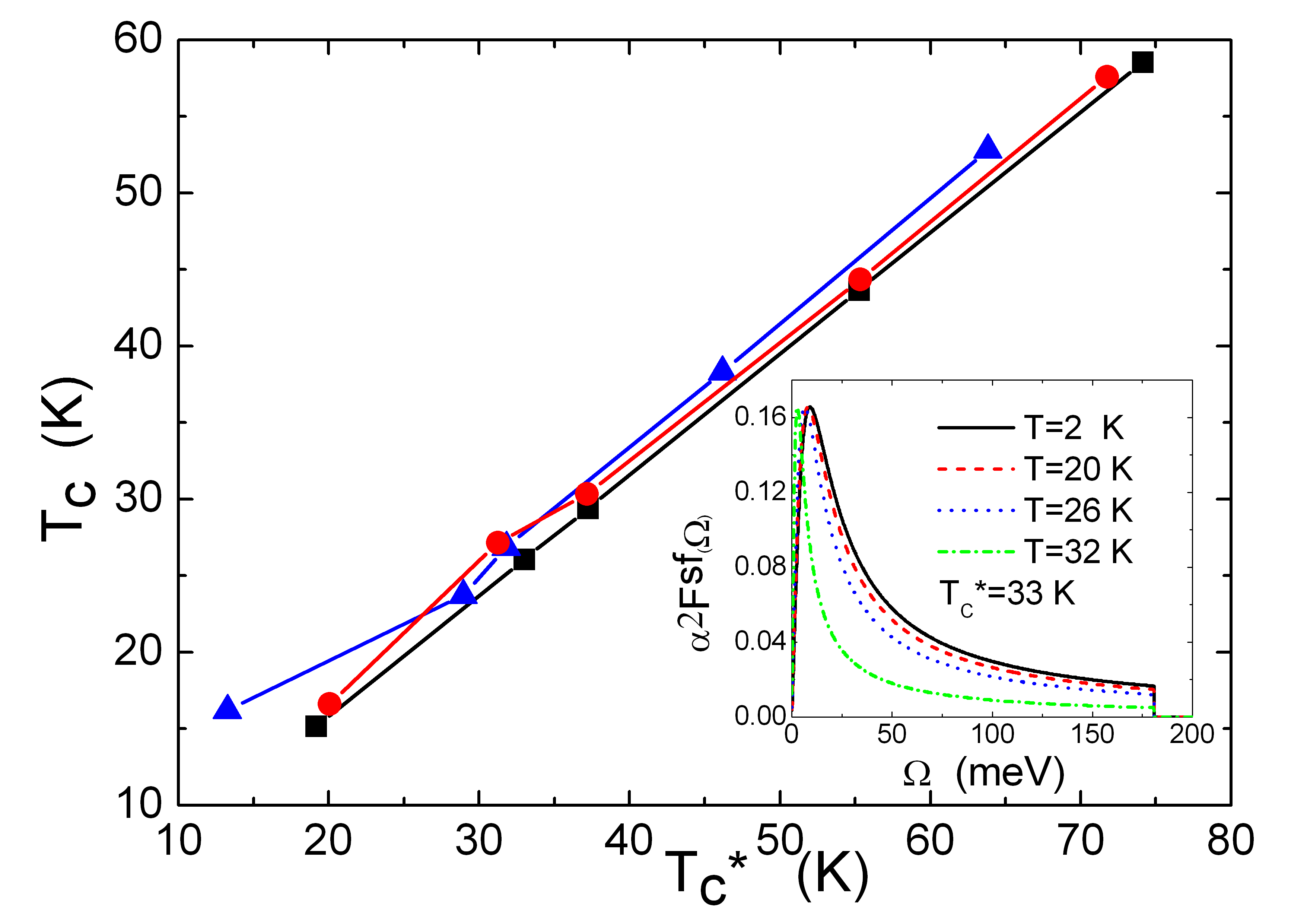}
\vspace{-5mm} \caption{(Color online) The calculated critical
temperature $T_{c}$ with \emph{feedback effect} versus standard
critical temperature $T^{*}_{c}$ in three different situations: only
interband \textit{sf} coupling with standard spectral functions
(black squares), interband \textit{sf} coupling with standard
spectral functions and small intraband \textit{ph} coupling (red
circles) and only interband \textit{sf} coupling with Lorentz
spectral functions (dark blue triangles). In the bottom right insert
the \textit{sf} spectral function, for the
$\mathrm{Ba(Fe_{x}Co_{1-x})_{2}As_{2}}$, at different temperatures
($T<T_{c}^{*}$) with the \emph{feedback effect}.}\label{Fig1}
\end{center}
\vspace{-5mm}
\end{figure}
\begin{figure}[!]
\begin{center}
\includegraphics[keepaspectratio, width=\columnwidth]{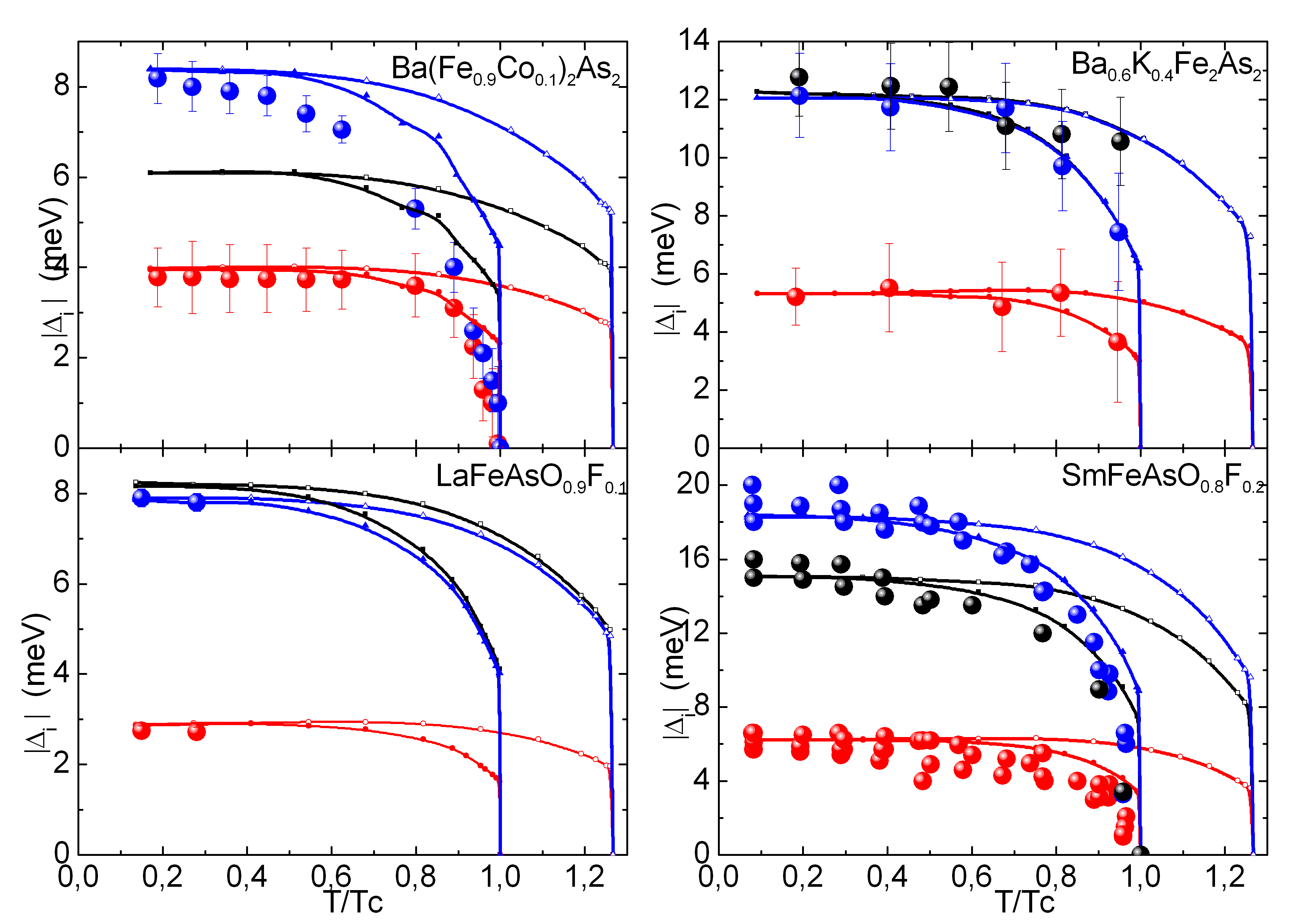}
\vspace{-5mm} \caption{(Color online) The calculated temperature
dependence of $|\Delta_{i}|$ from the solution of real axis
Eliashberg equations in the standard case (open symbol) and when the
\emph{feedback effect} is present (solid symbol): $|\Delta_{1}|$
black squares, $|\Delta_{2}|$ red circles and $|\Delta_{3}|$ dark
blue triangles. The experimental data
\cite{Gonnelli_La,Ding_ARPES_BaKFeAs,Daghero_Sm,Tortello2010} are
shown as big solid circles.}\label{Fig2}
\end{center}
\vspace{-5mm}
\end{figure}
I initially solve the imaginary-axis Eliashberg equations (eqs.
\ref{eq:EE1} and \ref{eq:EE2}) to calculate the low-temperature
value of the gaps (which are actually obtained by analytical
continuation to the real axis by using the technique of the Pad\'{e}
approximants) and so I fix the two free parameters of the model:
$\lambda_{13}$ and $\lambda_{23}$ ($\lambda_{12}$). By properly
selecting the values of $\lambda_{13}$ and $\lambda_{23}$
($\lambda_{12}$) it is relatively easy to obtain the experimental
values of the gaps with reasonable values of
$\lambda_{tot}=\frac{\sum_{ij}N_{i}(0)\lambda_{ij}}{\sum_{ij}N_{i}(0)}$
(between 1.72 and 2.04). However, in all the materials examined, the
high $2\Delta_{1,3} / k_B T_c$ ratio (of the order of 8-9) makes it
possible to reproduce \emph{also} the values of the large gap(s)
only if the calculated critical temperature $T^{*}_{c}$ is
considerably higher than the experimental one. For solving this
problem also present in the HTCS, I assume that exist a effect of
feedback \cite{Chubukov,FBB,FB} of the condensate and, in a
phenomenological way, I introduce in the Eliashberg equation a
temperature dependence of the representative boson energy
$\Omega_{0}(T)=\Omega_{0}tanh(1.76\sqrt{T^{*}_{c}/T-1})$ that
reproduces both the approximate gap temperature dependence in the
strong coupling case \cite{FB} and the experimental spin resonance
one \cite{Inosov}. The primary effect of this assumption is lowering
the critical temperature leaving unchanged the gap values at
$T<<T^{*}_{c}$ because the critical temperature is roughly
proportional to electron boson coupling constant and to
representative boson energy $\Omega_{0}(T)$ of the material: in this
case $\Omega_{0}(T)$ decreases and so $T_{c}$. For a completely
consistent procedure it should used
$\Omega_{0}(T)=\Omega_{0}\eta(T)$ where $\eta(T)$ is the temperature
dependence part of the superfluid density $\rho(T)=\rho(0)\eta(T)$
and $\rho(0)$ is the superfluid density at $T=0$ K. $\eta(T)$ is a
function of $\Delta_{i}(i\omega_{n})$ and so, in this way, the
numerical solution of Eliashberg equations become remarkably more
complex and time consuming. I am conscious that the temperature
dependence of $\Omega_{0}(T)$ is added ad hoc and it is not obtained
self-consistent but this is an attempt in order to determine if the
chosen path can lead to interesting results. What is important is
that this mechanism of \emph{feedback} can justify the experimental
values for the gaps, their dependence on temperature and the
critical temperature with a model that has only two free parameters.
Moreover, the parameters determined are reasonable and
$\lambda_{tot}$ is very similar for all four materials examined and
in agreement with the values proposed by other authors
\cite{dolghi}.
I solve the Eliashberg equations in three different situations: 1)
only \textit{sf} interband coupling is present and the \textit{sf}
spectral functions have usual shape; 2) \textit{sf} interband
coupling with a small \textit{ph} intraband contribution are present
and \textit{sf} spectral functions have usual shape; 3) only
\textit{sf} interband coupling is present and the \textit{sf}
spectral functions have Lorentz shape. In the first case the
coupling constant $\lambda_{tot}$ is in the range 1.72-2.04. The
results are almost independent from $\Omega_{max}$ because, for
example in the case of $BaFeCoAs$, multiplying $\Omega_{max}$ by a
factor two, I obtain the same values of the gaps and $T_{c}$ with
$\lambda_{tot}=1.68$ i.e. with a reduction of 0.18 which is very
small. The agreement with the experimental critical temperature is
good. It is noticeable the small variation of the total coupling in
the four compounds considered. In the second case there is also a
intraband phonon contribution, equal in any band and in any compound
for simplicity, with $\lambda^{ph}_{ii}=0.35$ and
$\Omega^{ph}_{0}=18$ meV that are the upper limits for the
\textit{ph} coupling constants and the representative \textit{ph}
energies \cite{Boeri2}. The \textit{ph} spectral functions have
Lorentzian shape \cite{Umma1} with the peaks at the same energy:
$\Omega_{ij}=\Omega^{ph}_{0}$and with half width always equal to 2
meV ( $\omega_{c}=12\Omega^{ph}_{0}$). $\lambda_{tot}$ and $T_{c}$
are practically the same as the previous case. This last fact
indicate that the effect of intraband phonon contribution is
negligible. In the third case (Lorentz shape of \textit{sf} spectral
functions) the agreement with the experimental critical temperatures
is very good in all compounds but the total coupling is more large
($2.38\leq\lambda_{tot}\leq 2.84$). In Fig. 1 it is possible to see
the linear relation between $T_{c}$ and $T^{*}_{c}$ in all three
examined cases. In table 1 are shown the inputs parameters of the
Eliashberg equations in the first and third case examined for the
four compounds. In table 2 are shown the calculated values of the
gaps and the critical temperatures $T_{c}$ and $T^{*}_{c}$ obtained
by numerical solution of Eliashberg equations.
Once the values of the low-temperature gaps were obtained, I
calculated their temperature dependence by directly solving the
three-band Eliashberg equations in the real-axis formulation instead
of using the analytical continuation to the real axis of the
imaginary-axis solution. Of course, the results of the two
procedures are virtually identical at low temperature. In all cases,
their behavior is rather unusual and completely different from the
BCS one, since the gaps slightly decrease with increasing
temperature until they suddenly drop close to $T_{c}$. This arises
from a complex non-linear dependence of the $\Delta$ vs. $T$ curves
on $\lambda_{ij}$ and is possible only in a strong-coupling regime
\cite{UmmaT}. Curiously in all four compounds the rate
$T^{*}_{c}/T_{c}$ is $1.27$. As it is shown in Fig. 2 the calculated
temperature dependencies of $|\Delta_{i}|$ are compared with the
experimental data and the agreement is very good (in the case of
$\mathrm{Ba(Fe_{x}Co_{1-x})_{2}As_{2}}$ I compare the temperature
dependence of the gaps with these particular experimental
values\cite{Tortello2010}: $\Delta_{1}=3.8$ meV and $\Delta_{2}=8.2$
meV and I find $\lambda_{12}=0.77$, $\lambda_{13}=1.05$ and
$T_{c}=23.43$ K).
In conclusion, I have shown that a simple Eliashberg three-band
model, with antiferromagnetic spin fluctuactions-electrons coupling,
in moderate strong-coupling regime with only two free parameters and
a \emph{feedback effect} can reproduce, in a \emph{quantitative}
way, the experimental critical temperature and the amplitude of the
energy gaps.

I thank R.S. Gonnelli, E. Cappelluti, Lara Benfatto and Sara Galasso
for the useful and clarifying discussions.

\end{document}